\documentstyle[11pt]{article}
\normalbaselines
\begin{document}
\bibliographystyle{unsrt}

\newtheorem{theorem}{Theorem}
\newtheorem{lemma}{Lemma}
\newtheorem{proposition}{Proposition}

\def\bea*{\begin{eqnarray*}}
\def\eea*{\end{eqnarray*}}
\def\ba{\begin{array}}
\def\ea{\end{array}}
\count1=1
\def\be{\ifnum \count1=0 $$ \else \begin{equation}\fi}
\def\ee{\ifnum\count1=0 $$ \else \end{equation}\fi}
\def\ele(#1){\ifnum\count1=0 \eqno({\bf #1}) $$ \else \label{#1}\end{equation}\fi}
\def\req(#1){\ifnum\count1=0 {\bf #1}\else \ref{#1}\fi}
\def\bea(#1){\ifnum \count1=0   $$ \begin{array}{#1}
\else \begin{equation} \begin{array}{#1} \fi}
\def\eea{\ifnum \count1=0 \end{array} $$
\else  \end{array}\end{equation}\fi}
\def\elea(#1){\ifnum \count1=0 \end{array}\label{#1}\eqno({\bf #1}) $$
\else\end{array}\label{#1}\end{equation}\fi}
\def\cit(#1){
\ifnum\count1=0 {\bf #1} \cite{#1} \else 
\cite{#1}\fi}
\def\bibit(#1){\ifnum\count1=0 \bibitem{#1} [#1    ] \else \bibitem{#1}\fi}
\def\ds{\displaystyle}
\def\hb{\hfill\break}
\def\comment#1{\hb {***** {\em #1} *****}\hb }

\newcommand{\TZ}{\hbox{\bf T}}
\newcommand{\MZ}{\hbox{\bf M}}
\newcommand{\ZZ}{\hbox{\bf Z}}
\newcommand{\NZ}{\hbox{\bf N}}
\newcommand{\RZ}{\hbox{\bf R}}
\newcommand{\CZ}{\,\hbox{\bf C}}
\newcommand{\PZ}{\hbox{\bf P}}
\newcommand{\QZ}{\hbox{\bf Q}}
\newcommand{\HZ}{\hbox{\bf H}}
\newcommand{\EZ}{\hbox{\bf E}}
\newcommand{\GZ}{\,\hbox{\bf G}}

\font\germ=eufm10
\def\goth#1{\hbox{\germ #1}}
\vbox{\vspace{38mm}}

\begin{center}
{\LARGE \bf Chiral Potts Rapidity Curve Descended from \\[2mm] 
Six-vertex Model and Symmetry Group of Rapidities}\\[10 mm] 
Shi-shyr Roan \\
{\it Institute of Mathematics \\
Academia Sinica \\  Taipei , Taiwan \\
(email: maroan@gate.sinica.edu.tw ) } \\[30mm]
\end{center}

\begin{abstract}
In this paper, we present a systematical account of the descending procedure from six-vertex model to the $N$-state chiral Potts model through fusion relations of $\tau^{(j)}$-operators, following the works of Bazhanov-Stroganov and Baxter-Bazhanov-Perk. A careful analysis of the descending process leads to appearance of the high genus curve as rapidities' constraint for the chiral Potts models. Full symmetries of the rapidity curve are identified, so is its symmetry group structure. By normalized transfer matrices of the chiral Potts model, the $\tau^{(2)}T$ relation can be reduced to functional equations over a  hyperelliptic curves associated to rapidities, by which the degeneracy of  $\tau^{(2)}$-eigenvalues is revealed in the case of superintegrable chiral Potts model. 
\end{abstract}
\par \vspace{5mm} \noindent
{\it 2000 MSC}: 14H45, 14Q05, 82B23  \par \noindent
{\it 1999 PACS}:  05.50.+q, 02.30.Gp \par \noindent
{\it Key words}: Chiral Potts model, Six-vertex model, $\tau^{(2)}T$ relation, $T\widehat{T}$ relations, $\tau^{(j)}$ fusion relations. 

\vfill
\eject

\section{Introduction}
The purpose of this article is the revisit of known facts on the $N$-state chiral Potts model as a descendant of  six-vertex model, and functional relations in the chiral Potts model. The discussion will be mainly based on two notable papers \cite{BBP, BazS} in this recently-discovered solvable lattice model, (for "descendants" of a more general class of vertex models, see a recent work of R. Baxter \cite{B04}). The formulae appeared in this work are to a large extent borrowed from \cite{BBP}, and extensive uses are also made of other  Baxter's works. Hence, the present article lays no claim to deep originality. In the way, our motivation is an attempt of better understanding the significance behind many identities in Baxter's papers, and clarifying the mathematical content of formulae appeared in \cite{BBP}. However, after the analysis made on the descending procedure, our effort leads to the appearance of chiral Potts rapidity's constraint in a natural way from the viewpoint of descendant of six-vertex model. Afterwards, we proceed to determine all symmetries of the rapidity curve, of which the large finite symmetry group structure has been widely believed for its role in solvability of the model; we further explore the degenerate eigenvalues of the six-vertex model through the chiral Potts transfer matrices, as an analogy to the discussion in \cite{FM} for eight-vertex model for the root of unity cases. We therefore hope that the reader will still find our presentation to be of independent interest.

In the study of two-dimensional solvable $N$-state chiral Potts model (for a brief
history account, see, e.g. \cite{Mc} section 4.1 and references therein), 
"rapidities"  of the statistical model are described by elements $[a, b, c, d]$ in the projective 3-space $ \PZ^3$ satisfying the following equivalent sets of equations,
\begin{eqnarray}
{\goth W}  : \, \  \left\{ \begin{array}{l}
ka^N + k'c^N = d^N , \\
kb^N + k'd^N = c^N . \end{array}  \right.   &
\Longleftrightarrow  &
\left\{ \begin{array}{l}
a^N + k'b^N = k d^N  , \\
k'a^N + b^N =
k c^N , \end{array}  \right.
\label{rapidC}
\end{eqnarray}
where $k, k'$ are parameters with $k^2 + k'^2 = 1$, and $k' \neq \pm 1, 0$. The above relations  define ${\goth W}$ as an algebraic curve of genus $N^3-2N^2+1$, which will be called the rapidity curve throughout this paper. 
For simplicity, we shall confine our discussion of chiral Potts models only on the full homogeneous lattice by taking $p=p'$ in \cite{BBP}. Note that a generalized column-inhomogeneous $\tau_2(t_q)$ model and its corresponding row-to-row transfer matrix functional relations without the conditions (\ref{rapidC}) are recently discussed by Baxter in \cite{B049}. 

It is known in \cite{BazS} that when descending the six-vertex model to $N$-state chiral Potts model, one first solves the Yang-Baxter $RLL$-relation of the six-vertex model to obtain the $L$-solution with operator-entries acting on the "quantum space" $\CZ^N$ (in the terminology of quantum inverse scattering method, see, e.g. \cite{F, KS}), and with  parameters depending on an arbitrary four-vector ratio $p =[a, b, c, d] \in \PZ^3$, (for the explicit form, see formulas (\req(G)) (\req(YBe)) of this paper). The trace of $L$-operator gives rise to a commuting family of operators $\tau^{(2)}_p(t)$ for $t \in \CZ$.  
Our attempt is to search certain principles which impose the rapidities' constraint (\ref{rapidC}) for the chiral Potts model through descending processes from the six-vertex model. By examining the formulas of functional relations involved in the chiral Potts transfer matrices $T_p(q)$ in \cite{BBP}, we observe that the fusion relations of $\tau^{(j)}_p$-operators, which are induced from $\tau^{(2)}_p$, are equivalent to the rapidity constraint (\ref{rapidC}) on $p$. Then we go on to clarify all symmetries of the curve (\ref{rapidC}), and identify the structure of the automorphism group ${\rm Aut}({\goth W})$ through three hyperelliptic curves of genus $(N-1)$ associated to ${\goth W}$. One such hyperelliptic curve is defined by the variables $t= \frac{ab}{cd}, \lambda=\frac{d^N}{c^N}$ with the relation
\be
W_{k'} : \ \ t^N = \frac{(1- k' \lambda )( 1 - k' \lambda^{-1}) }{k^2 } \ .
\ele(Wk')
By normalizing the transfer matrices $T_p(q)$ as in \cite{B90}, the $T\widehat{T}$ and $\tau^{(2)}T$ relations on ${\goth W}$ can be reduced to functional equations of operators on $W_{k'}$. Then by using the $(t, \lambda)$-variable form of $\tau^{(2)}T$ relation derived in this paper, (which to the best of our knowledge, has not previously appeared in the literature), we are able to show that the degeneracy of  $\tau^{(2)}_p$-eigenvalues appears when $p$ is the superintegrable element, an analogy to the discussion of $T$-$Q_{72}$ relation for the eight-vertex model in \cite{FM}.

The remainder of this article is organized as follows. In Sec. 2, we begin our discussion by briefly reviewing the rapidities and Boltzmann weights of the $N$-state chiral Potts model in literatures (e.g., \cite{AMPT, BPA}). In Sec. 3, we start with a solution $\tau^{(2)}_p$ of the Yang-Baxter equation of six-vertex model with the parameter $p \in \PZ^3$ in \cite{BazS}, then define $\tau^{(j)}_p$-operators for $0 \leq j \leq N$ through fusion relations, originally appeared in the study of chiral Potts models in \cite{BBP}. A careful analysis of the fusion relations of $\tau^{(j)}_p$, the constraint of $p$ in rapidity curve ${\goth W}$ naturally arises as an consequence of these relations. 
In Sec. 4 we briefly review of results in \cite{BBP} about the chiral Potts transfer matrices $T_p(q)$ for $p, q \in {\goth W}$, $\tau^{(2)}T$ and $T\widehat{T}$ relations, and their connections with $\tau^{(j)}_p$. In this way, our $\tau^{(j)}$-fusion approach to the rapidity curve in the previous section can be better understood by the original source we are based upon. 
In Sec. 5, we first recall the relation of the rapidity curve of $N$-state chiral Potts model and three hyperelliptic curves of genus $(N-1)$ with $D_N$-symmetry in \cite{B91, R92}. Through this, we determine the full symmetries of the rapidity curve, and its group structure. In particular, the symmetries already appeared in literatures (e.g., in \cite{AP, BPA, B2}) do exhaust all the symmetries of  rapidities of the $N$-state chiral Potts model, and the order of the symmetry group ${\rm Aut}({\goth W})$ is equal to $4N^3$. In Sec. 6, by using the normalized operator $V_p (t_q, \lambda_q)$ of $T_p(q)$ as in \cite{B90}, we reduce the $\tau^{(2)}T$ and $T\widehat{T}$ relations from ${\goth W}$ to functional equations on the hyperelliptic curve $W_{k'}$ in (\req(Wk')). Note that the $(t, \lambda)$-form of $T\widehat{T}$ relations was previously used in the effective discussions of "ground state" energy \cite{B90} and the excitation spectrum \cite{MR}, both in the thermodynamic limit of an infinite lattice. Using the $(t, \lambda)$-form of $\tau^{(2)}T$ relation, one can see the degenerate $\tau^{(2)}_p$-eigenvalues  from the $V_p$-eigenvalues in the superintegrable case.  We close in Sec. 7 with some concluding remarks.

{\bf Notations.}
To present 
our work, we prepare some notations. In this
paper, 
$\ZZ, \RZ, \CZ$ will denote 
the ring of integers, real, complex numbers
respectively, $\ZZ_N=
\ZZ/N\ZZ$,  and ${\rm i} = \sqrt{-1}$. For $N \geq 2$, we fix the $N^{\rm th}$ root of unity, $$\omega= e^{\frac{2 \pi i}{N}} ,
$$
and $\CZ^N $ is the vector space consisting of all $N$-cyclic vectors with the basis $\{ | n \rangle \}_{ n \in \ZZ_N}$.
For a positive integers $n$, we denote
by $\stackrel{n}{\otimes} \CZ^N$ the tensor
product of $n$-copies of the vector space $\CZ^N$.

\section{The Rapidity Curve of $N$-state Chiral Potts Model}
Let  $X, Z$  be the 
operators of $\CZ^N $ defined by $X |n \rangle = | n +1 \rangle$, $ Z |n \rangle = \omega^n |n \rangle $ for $n \in \ZZ_N$. Then $X, Z$ satisfy the Weyl  
relation and $N^{\rm th}$-power identity property : $XZ= \omega^{-1}ZX$, $X^N=Z^N=1$. 

We shall denote the four-vector ratio by $[a, b, c, d]$ for  a non-zero vector $(a, b, c, d) \in \CZ^4 $. The collection of all four-vector ratios is the projective $3$-space $\PZ^3$. Hereafter we shall always use the variables $x, y, \mu$ to denote the following component-ratios of an element $[a, b, c, d] \in \PZ^3$,
\be
x := \frac{a}{d} , \ \ \  y :=  \frac{b}{c} , \ \ \ \mu : =   \frac{d}{c} \ .
\ele(xym)
Then $x, y, \mu$ can be considered as affine coordinates of $\PZ^3$. 
From now on, we shall denote elements in $\PZ^3$ simply by $p, q, r, \cdots$ etc. The coordinates of an element, say $p$, will be written in the forms $a_p, b_p, x_p, \cdots $ so on, whenever if it will be necessary to specify the element $p$. 

It is known that rapidities of the $N$-state chiral Potts model form the projective curve ${\goth W}$ (\ref{rapidC}) in $\PZ^3$.  In terms of the affine coordinates $(x , y , \mu )$ in (\req(xym)), an equivalent form of defining equations for ${\goth W}$ is given by
\be
k x^N  = 1 -  k'\mu^{-N}, \ \ \  k y^N  = 1 -  k'\mu^N \ , \ \ (x , y , \mu ) \in \CZ^3 \ .
\ele(xymu)
Define 
$$
e^{{\rm i} \theta_p} = e^{\frac{-\pi {\rm i}}{N}} y_p \ , \ e^{{\rm i} \phi_p} = x_p \ , \ \ 
u_p = \frac{N( \theta_p + \phi_p )}{2} \ , \ \ \ v_p = \frac{N(\theta_p - \phi_p)}{2} \ .
$$
By eliminating the valuable $\mu^N$ in (\req(xymu)) , ${\goth W}$ becomes a $N$-fold unramified cover of the genus $(N-1)^2$ curve,
\be
x^N + y^N = k ( 1 + x^N y^N )  ,  \ \  ( {\rm equivalently },  \ \ \sin v_p = k \sin u_p \ ) \ .
\ele(xy) 
 
The Boltzmann weights
$W_{p,q},\overline{W}_{p,q}$ of the $N$-state chiral Potts model are defined by the  coordinates of $p, q \in {\goth W}$ with the expressions: 
\begin{eqnarray*}
\frac{W_{p,q}(n)}{W_{p,q}(0)}  = \prod_{j=1}^n
\frac{d_pb_q-a_pc_q\omega^j}{b_pd_q-c_pa_q\omega^j} &\bigg( = (\frac{\mu_p}{\mu_q})^n \prod_{j=1}^n
\frac{y_q-\omega^j x_p}{y_p- \omega^j x_q } = (\frac{\cos N(\theta_q - \phi_p)/2}{\cos N(\theta_p - 
\phi_q)/2})^{\frac{-n}{N}}\prod_{j=1}^n \frac{\sin ( \frac{-\theta_q + \phi_p}{2} + \frac{\pi(2j-1)}{2N} )}{\sin ( \frac{- \theta_p + 
\phi_q}{2} + \frac{\pi(2j-1)}{2N} )} \bigg) 
\ , \\
\frac{\overline{W}_{p,q}(n)}{\overline{W}_{p,q}(0)} 
 = \prod_{j=1}^n
\frac{\omega a_pd_q-
d_pa_q\omega^j}{ c_pb_q- b_pc_q \omega^j} &\bigg( = ( \mu_p\mu_q)^n \prod_{j=1}^n \frac{\omega x_p - \omega^j x_q }{ y_q- \omega^j y_p } = (\frac{\sin N(\phi_q - \phi_p)/2}{\sin N(\theta_p - \theta_q)/2})^{\frac{- n}{N}}\prod_{j=1}^n \frac{\sin ( \frac{\phi_q - \phi_p}{2} + \frac{\pi(j-1)}{N} )}{\sin ( \frac{\theta_p - \theta_q }{2} + \frac{\pi(j-1)}{N} )} \bigg) \ 
\end{eqnarray*}
(see, e.g. \cite{AMP}).
By the rapidities' constraint (\ref{rapidC}), the above Boltzmann weights have the $N$-periodicity property for $n$. Equivalently to say, Boltzmann weights are represented by two cyclic vectors, $( W_{p,q}(n))_{n \in \ZZ_N}$ and $(\overline{W}_{p,q}(n))_{n \in \ZZ_N}$, of $\CZ^N$ with the ratio-conditions: $
\frac{W_{p,q}(n)}{W_{p,q}(n-1)}  = 
\frac{d_pb_q-a_pc_q\omega^n}{b_pd_q-c_pa_q\omega^n}$ , $\frac{\overline{W}_{p,q}(n)}{\overline{W}_{p,q}(n-1)} 
= \frac{\omega a_pd_q-d_pa_q\omega^n}{ c_pb_q- b_pc_q \omega^n}$. For convenience, we shall assume $W_{p, q}(0) = \overline{W}_{p,q}(0) = 1$ without loss of generality.

\section{Six-vertex Model and Fusion Relations of $\tau^{j}$'s}
By a remarkable paper \cite{BazS}, Bazhanov and Stroganov found that for each element $[a, b, c, d] \in \PZ^3$, there associates a solution $G(t)$ of Yang-Baxter (YB) relation for the six-vertex model in terms of the operators $X, Z$. In the terminology of quantum inverse scattering method, $G (t)$ is put into the following 2-by-2 matrix form with operator-entries acting on "quantum space" $\CZ^N$, 
\bea(l)
 b^2 G (t)  = b^2 \left( \begin{array}{cc}
       G_{0, 0}  & G_{0, 1}  \\
       G_{1, 0} & G_{1, 1}   
\end{array} \right) = \left( \begin{array}{cc}
       b^2 - t d^2 X & ( bc - \omega a d X )  Z   \\
       -t (b c - a d X ) Z^{-1} & - t c^2  + \omega  a^2  X  
\end{array} \right)  \ , \ \ \
t \in \CZ \ ,
\elea(G)
and satisfies the YB relation:
\be
R(t/t') (G (t) \bigotimes_{aux}1) ( 1
\bigotimes_{aux} G(t')) = (1
\bigotimes_{aux} G(t'))(G (t)
\bigotimes_{aux} 1) R(t/t') \ ,
\ele(YBe)
where $R(t)$ is the following matrix\footnote{The YB-relation solution $G(t)$ we describe here is in accordance with discussions in Sec. 4 of \cite{BBP}, which will be briefly reviewed in Sec. 4 of the present article. Consequently, the R-matrix $R(x)$ in our content is required to vary the form appeared as (2.1) in \cite{BazS}. Indeed, its entries are the Boltzmann weights of six-vertex model described in (5) of \cite{B049}, also previously discussed  in \cite{PS}.} of 2-tensor of 
"auxiliary  space" $\CZ^2$,
$$
R(t ) = \left( \begin{array}{cccc}
        t \omega - 1  & 0 & 0 & 0 \\
        0 &t-1 & \omega  - 1 &  0 \\ 
        0 & t(\omega  - 1) &\omega( t-1) & 0 \\
     0 & 0 &0 & t \omega - 1    
\end{array} \right) \ .
$$
By the auxiliary-space matrix-product and quantum-space tensor-product, the operator of a finite size $L$, 
\be
\bigotimes_{j=1}^L G_j (t)  = G_1 (t) \bigotimes \cdots \bigotimes G_L (t) \ , \ \ \ \ 
G_j (t) :=  G (t)  \ ,
\ele(Gj) 
again satisfies the YB relation (\req(YBe)), hence the traces, $\bigg\{ {\rm tr}_{aux} ( \bigotimes_{j=1}^L G_j (t) ) \bigg\}_{t \in \CZ}$, form a
family  of commuting operators of  $\stackrel{L}{\otimes} \CZ^N$. As $G_j(t)$ depends on the parameter $p =[a, b, c, d] \in \PZ^3$, it will be written by $G_{p, j}(t)$ as well. Define  
the $\tau^{(2)}_p$-operator by
\be
\tau^{(2)}_p (t) = {\rm tr}_{aux} ( \bigotimes_{j=1}^L G_{p, j}( \omega t)) \ \ \ \ {\rm for} \ \  t \in \CZ , 
\ele(tau2)
which again form a commuting family of operators acting on $\stackrel{L}{\otimes} \CZ^N$ for an arbitrary given $p \in \PZ^3$. The spin-shift operator of $\stackrel{L}{\otimes} \CZ^N$ will be denoted by $X \ (:=\bigotimes_{j=1}^L X_j)$, which has the eigenvalues $\omega^Q$ for $ Q \in \ZZ_N$. 
In the study of chiral Potts transfer matrices in \cite{BBP}, there are families of operators, $\tau^{(j)}_p$ for $ 0 \leq j \leq N$, constructed from  the $\tau^{(2)}_p$-family (\req(tau2))  by setting $\tau^{(0)}_p(t)= 0 $, $\tau^{(1)}_p(t)= I$, and the following "fusion relations" (see (4.27) of \cite{BBP} ):
\begin{eqnarray}
&\tau^{(j)}_p(t) \tau^{(2)}_p(\omega^{j-1} t) = z( \omega^{j-1} t ) X \tau^{(j-1)}_p(t) + \tau^{(j+1)}_p(t)  , & 1 \leq j \leq N \ , \label{Fus1} \\
& \tau^{(N+1)}_p(t):= z(t ) X \tau^{(N-1)}_p( \omega t) + u (t) I \ & \label{Fus2} 
\end{eqnarray}
with $z(t):= (\frac{\omega \mu_p^2 (x_p y_p - t )^2 }{y_p^4})^L $, $
u (t) := \alpha_p (\lambda) + \alpha_p (\lambda^{-1})$ where $t, \lambda$ are related by 
(\req(Wk')), and 
\be
\alpha_p ( \lambda ) = \bigg(\frac{k'(1-\lambda_p \lambda)^2}{ \lambda (1-k' \lambda_p)^2 }\bigg)^L \ \ (= \bigg(\frac{(y_p^N-x^N)(t_p^N-t^N)}{ y_p^{2N}(x_p^N-x^N) }\bigg)^L
\ \ {\rm if} \ p \in {\goth W} ) \ .
\ele(alpha) 
Note that $\alpha_p (\lambda) + \alpha_p (\lambda^{-1})$ can be expressed as a polynomial of $\lambda + \lambda^{-1}$, hence a polynomial of $t^N$.

By (\ref{Fus1}), one can express  $\tau^{(j)}_p (t)$ for $j > 2$ as a "polynomial" of $\tau^{(2)}_p$ of degree $(j-1)$ with coefficients in powers of $X$, e.g.,
$$
\begin{array}{ll}
\tau^{(3)}_p(t)  =& \tau^{(2)}_p(t)\tau^{(2)}_p(\omega t) - X z(\omega t ) , \\
\tau^{(4)}_p(t)  =& \tau^{(2)}_p(t)\tau^{(2)}_p(\omega t)\tau^{(2)}_p(\omega^2 t) - X z(\omega t )\tau^{(2)}_p(\omega^2 t) - X \tau^{(2)}_p(t) z(\omega^2 t ) , \\
\tau^{(5)}_p(t) =& \tau^{(2)}_p(t)\tau^{(2)}_p(\omega t)\tau^{(2)}_p(\omega^2 t)\tau^{(2)}_p(\omega^3 t) - X z(\omega t )\tau^{(2)}_p(\omega^2 t)\tau^{(2)}_p(\omega^3 t) - X \tau_p^{(2)}(t) z(\omega^2 t )\tau^{(2)}_p(\omega^3 t) \\
& - X z(\omega^3 t)\tau^{(2)}_p(t)\tau^{(2)}_p(\omega t) + X^2 z(\omega t ) z(\omega^3 t) . 
\end{array}
$$
Indeed, by induction argument one can show the following expressions of $\tau^{(j)}_p(t)$ for $ 2 \leq j \leq N+1 $ in terms of $\tau^{(2)}_p $ and $X$:
\begin{eqnarray}
\tau^{(j)}_p(t)  = \prod_{s=0}^{j-2} \tau^{(2)}_p(\omega^j t) + \sum_{k=1}^{[\frac{j-1}{2}] } (-X)^k  \sum_{1 \leq i_1 <'i_2 <' \cdots <' i_k \leq j-2}\prod_{\ell=1}^k \bigg( \frac{z(\omega^{i_\ell} t)}{\tau^{(2)}_p(\omega^{i_\ell-1 }t )\tau^{(2)}_p(\omega^{i_\ell }t)} \prod_{s=0}^{j-2} \tau^{(2)}_p(\omega^j t) \bigg) \  \label{tauF}
\end{eqnarray}
where the notion $i_\ell <' i_{\ell+1}$ means $i_\ell + 1 < i_{\ell+1}$.
Therefore, $\tau^{(j)}_p(t) $ commutes with $\tau^{(j')}_p(t') $ for all $j, j', t , t'$. 
However, (\ref{Fus2}) impose the constraint of $\tau^{(2)}_p (t)$, hence on $p$, of which the condition will be clear later on.

Using (\ref{tauF}), one obtains the following relations:
\begin{eqnarray}
\tau^{(N+1)}_p(t) =
\prod_{s =0}^{N-1} \tau^{(2)}_p(\omega^s t) + 
 \sum_{k=1}^{[\frac{N}{2}] } (-X)^k  \sum_{1 \leq i_1 <'i_2 <' \cdots <' i_k \leq N-1} \prod_{\ell=1}^k \bigg(  \frac{z(\omega^{i_\ell} t)}{\tau^{(2)}_p(\omega^{i_\ell-1 }t )\tau^{(2)}_p(\omega^{i_\ell }t)} \prod_{s=0}^{N-1} \tau^{(2)}_p(\omega^s t) \bigg) \nonumber ; \\
- z(t) X  \tau^{(N-1)}_p(\omega t)  = 
\sum_{k=1}^{[\frac{N}{2}] } z(t) (-X)^k  \sum_{0= i_1 <'i_2 <' \cdots <' i_k \leq N-2} \prod_{\ell=1}^k \bigg( \frac{z(\omega^{i_\ell} t)}{\tau^{(2)}_p(\omega^{i_\ell -1}t )\tau^{(2)}_p(\omega^{i_\ell }t)} \prod_{s=1}^{N-2} \tau^{(2)}_p(\omega^s t) \bigg) . \nonumber
\end{eqnarray}
By which, the relations (\ref{Fus1}) and (\ref{Fus2}) give rise to the functional equation of $\tau^{(2)}_p (t)$:
\begin{eqnarray}
 F_{\tau^{(2)}_p} ( t ) = u ( t ) I \ , \label{tau2eq}
\end{eqnarray}
where $F_{\tau^{(2)}_p} ( t )$ is defined by
\begin{eqnarray}
 F_{\tau^{(2)}_p} ( t ) : = \prod_{s =0}^{N-1} \tau^{(2)}_p(\omega^s t) + \sum_{k=1}^{[\frac{N}{2}] } (-X)^k  \sum_{ I_k  \in {\cal I}_k } \prod_{i \in I_k} \bigg( \frac{z(\omega^{i} t)}{\tau^{(2)}_p(\omega^{i-1 }t)\tau^{(2)}_p(\omega^{i }t)} \prod_{s=0}^{N-1} \tau^{(2)}_p(\omega^s t) \bigg),   \label{F2d}
\end{eqnarray}
with the index set ${\cal I}_k$ consisting of subsets $I_k$ of $\ZZ_N$ with $k$ distinct elements such that $i \not\equiv i'+1 \pmod{N}$ for all $i , i' \in I_k$. 
For examples, for $N=2,3, 4$, (\ref{F2d}) is given by
$$
\begin{array}{ll}
N=2, & F_{\tau^{(2)}_p} ( t ) =\tau^{(2)}_p(t) \tau^{(2)}_p(- t) -  (z(t )+z( - t )) X   ; \\
N=3, & F_{\tau^{(2)}_p} ( t ) = \prod_{j=0}^2\tau^{(2)}_p(\omega^j t) -  ( \sum_{j=0}^2 z(\omega^j t)\tau^{(2)}_p(\omega^{j+1} t) )  X \ ; \\
N=4, & F_{\tau^{(2)}_p} ( t )= \prod_{j=0}^3 \tau^{(2)}_p(\omega^j t) - ( \sum_{j=0}^3 z(\omega^j t)\tau^{(2)}_p(\omega^{j+1} t) )X  + (z(t) z(\omega^2 t) + z(\omega t)z(\omega^3 t) )X^2  .
\end{array}
$$
The functional equation (\ref{tau2eq}) for $\tau^{(2)}_p (t)$, equivalently the relations (\ref{Fus1}) and (\ref{Fus2}), naturally imposes the constraint on $p$, and it turns out to be the requirement of $p$ as an element in the rapidity curve ${\goth W}$. 
Indeed, we have the following characterization of ${\goth W}$.
\begin{theorem}\label{thm:tauW}
For $p \in \PZ^3$, the relation $(\ref{tau2eq})$ of $\tau^{(2)}_p (t)$ for $L=1$ is equivalent to that $p$ is an element of the rapidity curve ${\goth W}$.
\end{theorem}
{\it Proof.} We now consider the relation (\ref{tau2eq}) only for $L=1$. 
By (\req(G)) and the expression of $u(t)$ in (\ref{Fus2}), we have
\begin{eqnarray} 
&\tau^{(2)}_p (t) = y_p^{-2} \bigg(  \mu_p^2  ( \omega x_p^2  - \omega t   )  X + (y_p^2  - \omega t )I \bigg) ,  \ \ 
u(t) =  \frac{(1+k'^2)(1+ \mu_p^{2N} ) -4k' \mu_p^N}{(1-k' \mu_p^N)^2} -  \frac{(1-k'^2)(1+ \mu_p^{2N})}{(1-k' \mu_p^N)^2 }t^N . \ \ \ \ \label{L1} 
\end{eqnarray}
We are going to derive an explicit form of (\ref{F2d}) for $L=1$. By the invariant property under $t \mapsto \omega t$,  $\prod_{j=0}^{N-1} \tau^{(2)}_p (\omega^j t)$ is expressed by powers of $t^N$; so is 
$\sum_{ I_k  \in {\cal I}_k } \prod_{i \in I_k} ( \frac{z(\omega^{i} t)}{\tau^{(2)}_p(\omega^{i-1 }t)\tau^{(2)}_p(\omega^{i }t)} \prod_{s=0}^{N-1} \tau^{(2)}_p(\omega^s t) )$
as the index set ${\cal I}_k$ is invariant under the translation by $1 \pmod{N}$. It is known that $X^N = I$, and $I, X, \cdots, X^{N-1}$ form a set of linearly independent matrices.
By (\ref{L1}), one has 
\begin{eqnarray*}
\prod_{j=0}^{N-1} \tau^{(2)}_p (\omega^j t) = c_0(p, t) + c_N(p, t) + \sum_{j=1}^{N-1}  {N
 \choose j}  c_j(p, t) X^j ,  
\end{eqnarray*}
and 
\begin{eqnarray*}
(-X)^k \sum_{I_k \in {\cal I}_k}  \prod_{i \in I_k} \bigg( \frac{z(\omega^i t)}{\tau^{(2)}_p(\omega^{i-1 }t )\tau^{(2)}_p(\omega^i t)} \prod_{s=0}^{N-1} \tau^{(2)}_p(\omega^s t)\bigg)= (-1)^k |{\cal I}_k|   \sum_{j=k}^{N-k} {N-2k \choose j-k} c_j(p, t) X^j 
\end{eqnarray*}
for $1 \leq k \leq [\frac{N}{2}]$, where $c_j(p, t):= y_p^{-2N}\mu_p^{2j} ( \omega^j x_p^{2j}y_p^{2N-2j} - t^N )$ for $0 \leq j \leq N$. Hence $(\ref{F2d})_{L=1}$ has the following expression:
\begin{eqnarray}
F_{\tau_p^{(2)}}(t) = &( c_0(p, t) + c_N(p, t) ) + \sum_{j=1}^{[\frac{N}{2}]-1} \bigg( \sum_{k=0}^j (-1)^k | {\cal I}_k | {N-2k \choose j-k} \bigg) ( c_j(p, t) X^{j} + c_{N-j}(p, t) X^{N-j})  \nonumber \\
&+ 2^{N-2[\frac{N}{2}]-1} \bigg( \sum_{k=0}^{[\frac{N}{2}]} (-1)^k | {\cal I}_k | {N-2k \choose [\frac{N}{2}]-k} \bigg) ( c_{[\frac{N}{2}]}(p, t) X^{[\frac{N}{2}]} + c_{N-[\frac{N}{2}]}(p, t) X^{N-[\frac{N}{2}]}) . \ \ \label{cj}
\end{eqnarray}
Set $p= s := [ \sqrt{t}, y \mu^{-1}, \mu^{-1}, 1]$ in (\ref{cj}) with generic $t, y, \mu$. By (\ref{L1}) one has $\tau^{(2)}_s (t) = (1  - \omega y^{-2} t )I$, which implies $F_{\tau_s^{(2)}}(t)$ is a scalar operator, equivalently, the coefficients of $X^j$ for $1 \leq j \leq N-1$ in (\ref{cj}) are all equal zeros. As $c_j(s, t) \neq 0$ for $j \geq 1$, one obtains the following recurrence relations for $|{\cal I}_k|$'s: 
\be
 \sum_{k=0}^j (-1)^k | {\cal I}_k | {N-2k \choose j-k}  = 0 \ , \ \ \ \ \ j = 1, \ldots , [\frac{N}{2}] \ .
\ele(recIk)
By which,  (\ref{cj}) becomes the relation, $F_{\tau_p^{(2)}}(t) =  c_0(p, t) + c_N(p, t)$ for $p \in \PZ^3$.  
By (\ref{L1}) and the expressions of $c_0(p, t)$ and $c_N(p, t)$, the functional relation $(\ref{tau2eq})_{L=1}$ can be reduced to the following equation involving only with the scalar term:   
$$
y_p^{-2N} \bigg(y_p^{2N}+ \mu_p^{2N}   x_p^{2N}  - t^N( 1+ \mu_p^{2N} ) \bigg) = \frac{(1+k'^2)(1+ \mu_p^{2N} ) -4k' \mu_p^N}{(1-k' \mu_p^N)^2} -  \frac{(1-k'^2)(1+ \mu_p^{2N})}{(1-k' \mu_p^N)^2 }t^N .
$$
Then it is easy to see that the above relation is equivalent to the relations: $
k^2 y_p^{2N}  =  (1-k' \mu_p^N)^2$ and $ k^2 x_p^{2N}  = ( 1- k'\mu_p^{-N})^2$,
i.e.,  $p$ is an element of ${\goth W}$ by (\req(xymu)). 
$\Box$ \par  \noindent
{\bf Remark.} (1) As to the numerical values of $|{\cal I}_k|$'s, it is easy to see that  
$|{\cal I}_1|= N$ and $|{\cal I}_2|= \frac{N(N-3)}{2}$. However, for $k \geq 2$, it seems a non-trivial task to obtain  the explicit form  of $|{\cal I}_k|$ purely by the combinatoric method. The formula (\req(recIk)) provides a way to get the expression of $|{\cal I}_k|$ by recurrent relations. However, it would be interesting to have a certain combinatoric interpretation of (\req(recIk)).  

(2) The fusion relations (\ref{Fus1}) and (\ref{Fus2}) were originally derived from the chiral Potts model with $p \in {\goth W}$, which we will recall in the next section, hence the relation  (\ref{tau2eq}) holds for any site $L$ when $p \in {\goth W}$. 
$\Box$ \par \vspace{.2in} \noindent

\section{The $\tau^{(2)}T$ relation of Chiral Potts Model}
In this section, we recall the derivation of the fusion relations (\ref{Fus1}) (\ref{Fus2})  in the study of chiral Potts models in \cite{BBP}.  
For the $N$-state chiral Potts model on a lattice of the horizontal size $L$ with 
periodic boundary condition, the combined weights of
intersection between two consecutive rows give rise
to the transfer matrix acting on $\stackrel{L}{\otimes} \CZ^N$:
\begin{eqnarray}
T_p (q)_{\sigma, \sigma'} = \prod_{l=1}^L
\overline{W}_{p,q}(\sigma_l - \sigma_l')
W_{p,q}(\sigma_l - \sigma_{l+1}') \ , \ \ \ p, q \in {\goth W} \ ,
\label{Tpq}
\end{eqnarray}
where $\sigma= ( \sigma_1 , \ldots , \sigma_L) $ ,
$\sigma'= ( \sigma_1' , \ldots , \sigma_L') $ with $\sigma_l, \sigma_l' \in
\ZZ_N$. The Boltzmann
weights satisfy the
star-triangle relation:
$$
\sum_{d=0}^{N-1} \overline{W}_{qr}(b-d) W_{pr}(a-d) \overline{W}_{pq}(d-c) = R_{pqr} W_{pq}(a-b)  \overline{W}_{pr}(b-c) W_{qr}(a-c) \   
$$
with $R_{pqr}= \frac{f_{pq}f_{qr}}{f_{pr}}$ and $f_{pq}= \bigg( \frac{{\rm det}_N( \overline{W}_{pq}(i-j))}{\prod_{n=0}^{N-1} W_{pq}(n)}\bigg)^{\frac{1}{N}}$ \cite{BPA, MS}, which ensures the commutativity of  
transfer matrices for a fixed $p \in {\goth W} $:
$$
[ T_p (q) , \ \ T_p (q') ] = 0 \  \ , \ \ \ q , q' \in {\goth W} \ . 
$$
By (\req(xymu)), $T_p (q)$ depends only on the values of $(x_q, y_q)$, parametrized by the curve (\req(xy)) for $p$ fixed. Hence we shall also write $T_p(q)$ by $T_p (x_q, y_q)$ whenever it will be a convenient one. It is easy to see that $T_p(q)$ commutes with both the spin-shift operator $X$ of $\stackrel{L}{\otimes} \CZ^N$ and the spatial translation operator $S_R$, which takes the $j$th column to $(j+1)$th one for $1 \leq j \leq L$ with the 
identification $L+1 = 1$. We denote
$$
\widehat{T}_p(q) = T_p(q) S_R \ . 
$$

Now we describe the $\tau^{(2)}T$ relation in \cite{BBP, BazS}\footnote{$\tau^{(2)}_p ( q ), T_p (q)$ in this note are the operators $\tau^{(2)}_{k=0, q}$, $T_q$ in \cite{BBP} respectively.}.  By following the arguments in Sec. 4 of \cite{BBP}, one defines a 2-by-2 matrix, $G ( g', g ) = \bigg( G ( g', g )_{m, m'} \bigg)_{m, m'=0, 1}$, for two vectors $g = \sum_{k} g(k) | k \rangle , g' = \sum_{n} g'(n) | n \rangle \in \CZ^N$ by
$$
G (g', g ) = \sum_{n, k}  g'(n) g(k) G^k_n  \ , \ \ G^k_n := G ( |n \rangle , |k \rangle ) 
$$
where $G^k_n$ for $n, k \in \ZZ_N$ are given by (4.4), (3.37), (3.38) and (A.3) in \cite{BBP}: $G^k_n = 0$ except $k=n, n-1$, and 
$$
\begin{array}{ll}
{ G^n_n ~ }_{m, m'} & = (-1)^{m} \omega^{(m'-m)n +m }  (\frac{c_p}{b_p})^{m'+m}   t_q^{m} , \\
{ G^{n-1}_n ~ }_{m, m'} & = (-1)^{m-1} \omega^{(m'-m)(n-1)+1 }  (\frac{d_p}{b_p})^2(\frac{a_p}{d_p})^{m'+m}   t_q^{1-m'} \ ,
\end{array}
$$
where $t_q : = x_q y_q$. 
Hence ${G^k_n ~ }_{m,m'}$s can be put in form of a 2-by-2 matrix with the operator-valued acting on "quantum space" $\CZ^N$. Indeed, one has the following expression, 
$$
 \left( \begin{array}{cc}
       G_{0, 0}  & G_{0, 1}  \\
       G_{1, 0} & G_{1, 1}   
\end{array} \right) = G ( \omega t_q)
$$
where $G (t)$ is defined in (\req(G)). For convenience, we shall hereafter denote the following component-ratios of $p=[a, b, c, d] \in \PZ^3$ by
$$
t \ ( = t_p ) : = \frac{a b}{c d } = x y \ , \ \ \ \lambda \ (= \lambda_p) := \frac{d^N}{c^N} = \mu^N \ . 
$$
By (\req(xymu)), the variables $(t, \lambda) =(t_p, \lambda_p)$ for $p \in {\goth W}$ satisfy the relation (\req(Wk')), which defines a hyperelliptic curve of genus $N-1$. 
By (\req(tau2)), we have the commuting family $\tau^{(2)}_p (t_q)$ for an given $p \in {\goth W}$. The chiral Potts transfer matrices $T_p (q) $ constructed  from $\tau^{(2)}_p$-family in \cite{BBP, BazS} were along the line of "$T$-$Q$ relation" developed in \cite{B72-3}.  
Apply the ${\rm SL}_2$-gauge transform on the $j$th site $G_j (\omega t_q)$ in (\ref{Gj}) in the following manner: 
$$
H_j  = P_j^{-1} G_j (\omega t) P_{j+1} \ , \ \ \ P_j = \frac{1}{\sqrt{1+r_j^2}}\left( \begin{array}{cc}
        1  & r_j \\
     -r_j  & 1     
\end{array} \right) \ \ , \ ( P_{L+1} : = P_1 ) \ ,
$$
then the trace remains the same, i.e., $\tau^{(2)}_p (t_q) = {\rm tr}_{aux} ( \bigotimes_j H_j )$. The choice of the $P_j$'s above are made in searching some non-trivial kernel vector $g_j \in \CZ^N$ of $H_{j; 1,0}$ ( the left lower entry of $H_j $ ) for each $j$ so that one can construct a Bethe-equation-type eigenvalues of  $\tau^{(2)}_p (t_q)$. 
For $p, q \in {\goth W}$, one can solve of $r_j , g_j$ as follows: for each basis element $\beta = \otimes_{j} | \beta_j \rangle \in 
\stackrel{L}{\bigotimes} \CZ^N$ with $\beta_j \in \ZZ_N$, there associates a set of solutions $r_j$'s and the kernel vectors $g_j$'s, given by (4.14), (4.19 a) in \cite{BBP}:
$$
r^{\beta}_j = \omega^{1- \beta_{j-1}} x_q \ , \ \ \ \ \ 
g^{\beta}_j (n) = {y_p}^2 \overline{W}_{p, Uq}(n -\beta_{j-1}) W_{p, Uq} (n- \beta_j ) \ \ , \ \
 \beta_0 := \beta_L \ ,
$$
where $U$ is the following automorphism of ${\goth W}$,
$$
U : {\goth W} \longrightarrow {\goth W}, \  [a,b,c,d] \mapsto [\omega a,b,c,d] , \ \ \ \ ( (x, y, \mu) \mapsto (\omega x, y, \mu) ). 
$$
Furthermore, the vectors $g'^{\beta}_j=H_{j; 0,0}(g^{\beta}_j)$ and $g''^{\beta}_j=H_{j; 1,1}(g^{\beta}_j)$ have the expressions: 
$$
\begin{array}{l}
g'^{ \beta}_j (n) = \frac{(y_p -\omega x_q)(t_p -t_q)}{x_p - x_q} \overline{W}_{p, q}(n-\beta_{j-1}) W_{p, q} (n- \beta_j ) \ , \\
g''^{\beta}_j (n) = \frac{\omega \mu_p^2(x_p - \omega x_q)(t_p -\omega t_q)}{y_p - \omega^2 x_q}\overline{W}_{p, U^2q}(n -\beta_{j-1}-1) W_{p, U^2q} (n - \beta_j-1 ) \ ,
\end{array}
$$
(see (4.19  b, c) in \cite{BBP}). This implies
$$
\tau^{(2)}_p (t_q) ( \otimes_j g^{\beta}_j ) = \otimes_j g'^{\beta}_j + \otimes_j g''^{\beta }_j \ . 
$$
By the expression of $T_p (q)$ in (\ref{Tpq}), the above $g^{\beta}_j (n), g'^{\beta}_j (n)$ and $ g''^{\beta}_j (n)$ for all basis elements $\beta$  give rise to the following $\tau^{(2)}T$ relation (4.20) in \cite{BBP} (or (14) in \cite{B02}):
\be
\tau^{(2)}_p(t_q) T_p(\omega x_q, y_q) = \varphi_p(q) T_p(x_q, y_q) + \overline{\varphi}_p(Uq) X T_p(\omega^2x_q, y_q) \ , 
\ele(tauT)
where $
\varphi_p(q):= (\frac{(y_p -\omega x_q)(t_p -t_q)}{y_p^2(x_p - x_q)})^L$, $ \overline{\varphi}_p(q) := (\frac{\omega \mu_p^2(x_p -  x_q)(t_p - t_q)}{y_p^2(y_p - \omega x_q)})^L $. By which, one can express $\tau^{(2)}_p (t_q)$ in terms of $T_p$:
\be
\tau^{(2)}_p(t_q) = \bigg( \varphi_p(q) T_p(x_q, y_q) + \overline{\varphi}_p(Uq) X T_p(\omega^2x_q, y_q) \bigg) T_p(\omega x_q, y_q)^{-1} \ .
\ele(tau2T)
The commutativity of $T_p(q)$'s ensures that $\tau^{(2)}_p(t_q)$ commutes with $T_p( x_{q'}, y_{q'})$,
$$
[ \tau^{(2)}_p(t_q) , T_p( x_{q'}, y_{q'}) ] = 0 \ , \ \ \ \ {\rm for} \ \ p, q, q' \in {\goth W} \ .
$$
The  $\tau^{(j)}_p$'s in (\ref{Fus1}) (\ref{Fus2}) with $t=t_q$, are related to the transfer matrices $T_p(q)$ by the following $T\widehat{T}$ relations ((3.46) for $(l,k)=(j, 0)$ in \cite{BBP}, or (13) in \cite{B02} ):
\be
T_p(x_q, y_q) \widehat{T}_p( y_q, \omega^j x_q) =  r_{p, q} h_{j ; p, q} \bigg( \tau^{(j)}_p(t_q) + \frac{z(t_q)z(\omega t_q)\cdots z(\omega^{j-1} t_q)}{\alpha_p (\lambda_q)} X^j \tau_p^{(N-j)} (\omega^j t_q) \bigg) 
\ele(TThat)
for $0 \leq j \leq N$, where $r_{p, q}  = \bigg(\frac{N(x_p - x_q) (y_p - y_q) (t_p^N-t_q^N)}{(x_p^N - x_q^N) (y_p^N - y_q^N)(t_p - t_q) } \bigg)^L $ , $h_{j ; p, q} =  \bigg( \prod_{m=1}^{j-1} \frac{y_p^2 (x_p - \omega^m x_q)}{(y_p - \omega^m x_q)(t_p - \omega^m t_q) }\bigg)^L $and $\alpha_p (\lambda_q)$ is defined in (\req(alpha)). 
In particular, the relation (\req(TThat)) for $j=N$ becomes
\be
T_p(x_q, y_q) \widehat{T}_p( y_q,  x_q) =  \bigg( \frac{N y_p^{2N-2}(y_p - y_q)(y_p -  x_q) }{ (y_p^N - y_q^N)(y_p^N - x_q^N ) } \bigg)^L  \tau^{(N)}_p(t_q)    
\ele(TThatN)
(see (4.39) (4.44) in \cite{BBP}). Indeed, the operators $\tau^{(j)}_p$ were originally defined by the relation (\req(TThat)) in \cite{BBP}, and the fusion relations of $\tau^{(j)}$'s were derived from (\req(tauT)) and (\req(TThat)) with  the coefficient $z(t)$ in (\ref{Fus2}) satisfying $
z(t_q) =  \varphi_p(q) \overline{\varphi}_p(q)$. 
Using (\req(tau2T)) and (\ref{Fus1}), one can successively express  $\tau_p^{(j)} 
(q)$ for  $1 \leq j \leq N+1$ in terms of $T_p(q)$ ((4.34) in \cite{BBP}):
\bea(ll)
\tau^{(j)}_p(q) =&  T_p(x_q, y_q) T_p(\omega^j x_q , y_q) \sum_{m=0}^{j-1} \bigg( \varphi_p(q)\varphi_p(Uq) \cdots \varphi_p(U^{m-1}q)  \times \\
&\overline{\varphi}_p(U^{m+1}q) \cdots \overline{\varphi}_p(U^{j-1}q) 
T_p(\omega^m x_q, y_q)^{-1} T_p(\omega^{m+1}x_q, y_q)^{-1} X^{j-m-1} \bigg) \ .
\elea(taujT)
Then, by $\prod_{j=0}^{N-1} \varphi_p(U^j q) = \alpha_p (\lambda_q)$ and $\prod_{j=0}^{N-1} \overline{\varphi}_p(U^j q) = \alpha_p (\lambda_q^{-1})$, the relation  (\ref{Fus2}) automatically follows. 
In this way, one can interpret the expression of $\tau_p^{(2)}$ in (\req(tau2T)), or equivalently the $\tau^{(2)}T$ relation (\req(tauT)), as a $\tau^{(2)}_p$-solution of the functional equation (\ref{tau2eq}) when $p$ is an element of ${\goth W}$  .
By (\req(TThatN)) and the $T_p$-expression of $\tau^{(N)}_p(q)$ in (\req(taujT)), one obtains the functional equation of chiral Potts transfer matrices $T_p$ ((4.40) of \cite{BBP}):
$$
 \widehat{T}_p( y_q,  x_q) =    \sum_{m=0}^{N-1} C_{m; p}(q) T_p( x_q , y_q) T_p(\omega^m x_q, y_q)^{-1} T_p(\omega^{m+1}x_q, y_q)^{-1} X^{-m-1} \ , 
$$
where $C_{m; p}(q) =   \varphi_p(q)\varphi_p(Uq) \cdots \varphi_p(U^{m-1}q) 
\overline{\varphi}_p(U^{m+1}q) \cdots \overline{\varphi}_p(U^{N-1}q) ( \frac{N y_p^{2N-2}(y_p - y_q)(y_p -  x_q) }{ (y_p^N - y_q^N)(y_p^N - x_q^N ) })^L $.

\section{The Symmetry Group of Chiral Potts Rapidity Curve and its Relation with Hyperelliptic Curves with $D_N$-symmetry }
It is known that the rapidities of the chiral Potts model has a large finite symmetry group. In this section, we are going to identify the precise group structure of ${\rm Aut}({\goth W})$. As in \cite{B2}, we consider the following automorphisms of ${\goth W}$,
\bea(lll)
M^{(1)}&: [a,b, c, d] \mapsto [\omega a, b, c, \omega d ] , & (x, y, \mu) \mapsto (x , y , \omega \mu ) , \\
M^{(2)}&: [a,b, c, d] \mapsto [\omega a, \omega b, c,  d ] , & (x, y, \mu) \mapsto ( \omega x , \omega y , \mu ) , \\
M^{(3)}&: [a,b, c, d] \mapsto [ c, \omega^{\frac{1}{2}}d, \omega^{\frac{-1}{2}}a, \omega^{-1}b ] , & (x, y, \mu) \mapsto (\omega y^{-1} , \omega x^{-1} , \omega^{\frac{-1}{2}} x^{-1}y \mu^{-1} ) , \\
M^{(4)}&: [a,b, c, d] \mapsto [ a, b, \omega^{-1}c, d ] , &  (x, y, \mu) \mapsto ( x , \omega y , \omega \mu ) , \\
M^{(5)}&: [a,b, c, d] \mapsto [ d, \omega^{\frac{1}{2}}c, \omega^{\frac{-1}{2}}b, a ] , & (x, y, \mu) \mapsto (x^{-1} , \omega y^{-1} , \omega^{\frac{1}{2}} x y^{-1} \mu ) , \\
R &:  [a,b, c, d] \mapsto [b, \omega a, d, c ] , & (x, y, \mu) \mapsto ( y , \omega x , \mu^{-1} ) .
\elea(Sym)
Then one has
$$
R = M^{(3)} M^{(5)} = M^{(2)} M^{(5)} M^{(3)}, \, \ R^2 = {M^{(2)}},  \,\ {M^{(3)}}^2 = {M^{(5)}}^2 = {M^{(4)}}^N = 1 \ . 
$$
The diagonal symmetries of coordinates of ${\goth W}$ are expressed by
$$
\begin{array}{rl}
 U=M^{(1)} M^{(2)} {M^{(4)}}^{-1}:  [a,b, c, d] \mapsto [\omega a,b, c, d] ; \ \ \ &
{M^{(1)}}^{-1} M^{(4)} :  [a,b, c, d] \mapsto [ a,\omega b, c, d] ; \\
{M^{(4)}}^{-1} :  [a,b, c, d] \mapsto [ a, b, \omega c, d] ; \ \ \ &
{M^{(2)}}^{-1} M^{(4)} :  [a,b, c, d] \mapsto [a,b, c, \omega d] \ .
\end{array}
$$
Note that ${\goth W}/\langle M^{(1)} \rangle$ is represented by the curve (\req(xy)), of which for $N=3$ case, the Picard-Fuch equation of periods and algebraic geometry properties of theta function and the Jacobian variety  were investigated in details in \cite{D, DN, MS}.
Denote
$$
M^{(0)} = M^{(1)}M^{(2)}{M^{(4)}}^{-2} \ \ : [a, b, c, d] \mapsto [\omega a, b, \omega c,  d ] \ , \ \ \ {\rm equivalently} \ , \ \ (x, y , \mu ) \mapsto ( \omega x , \omega^{-1} y , \omega^{-1} \mu ) \ .
$$
Among the three morphisms $M^{(i)}$ for $i=0,1,2$, any two automorphisms generate a $\ZZ_N^2$-group acting freely on ${\goth W}$. Their quotient Riemann surfaces can be realized as members in the following one-parameter family of hyperelliptic curves:
$$
W_\kappa \ ( = W_{N, \kappa}) \ : \ \, \ T^N = \frac{(1 - \kappa
\Lambda)(1 - \kappa
\Lambda^{-1})}{1 - \kappa^2 } \ , \ \ \ (T,
\Lambda )
\in \CZ^2 \ ,
$$
where $\kappa$ is a complex parameter $\neq 0, \pm 1$.
For $N \geq 3$, the above family of curves is characterized by the  hyperelliptic curves of genus $N-1$ with $\ZZ_2 \times D_N$ symmetry group, where $D_N$ is the dihedral group (see,  Proposition 2 in \cite{R98}).
The hyperelliptic involution is given by
$$
\sigma : (T, \Lambda ) \mapsto (T, \Lambda^{-1})
\ ,
$$
and $D_N$ is generated by the automorphisms $\theta, \iota$ of order $N, 2$
respectively: 
$$
\theta:  (T, \Lambda ) \mapsto (\omega T,
\Lambda) \ , \ \, \ \, \
\iota: (T, \Lambda ) \mapsto (\frac{1}{T},
\frac{1- \kappa \Lambda}{\kappa - \Lambda} ) \ .
$$
It is known that the three $N^2$-unramified quotients of ${\goth W}$ can be realized as the following hyperelliptic curves ((25) in \cite{R92}):
\be
 W_{ k^\prime} \simeq {\goth W}/ \langle M^{(0)}, M^{(1)} \rangle  \ , \, \ \, \
W_{ {\rm i}k^\prime/k} \simeq {\goth W}/ \langle M^{(1)}, M^{(2)} \rangle \ , \, \ \, \ 
 W_{ k} \simeq {\goth W}/ \langle M^{(0)}, M^{(2)} \rangle  \ ,
\ele(CPHE)
with the coordinate-expression from ${\goth W}$ to hyperelliptic curves  given by 
\bea(ll)
{\goth W} \longrightarrow  W_{ k^\prime} , & [a, b, c, d] \mapsto (t , \lambda) = (\frac{ab}{cd}, \frac{d^N}{c^N}) ;  \\
{\goth W} \longrightarrow  W_{ {\rm i}k^\prime/k} , & [a, b, c, d] \mapsto (T_r , \Lambda_r) = (\frac{ac}{bd}, \frac{{\rm i} d^N}{ b^N} ) ; \\
{\goth W} \longrightarrow  W_{ k } , & [a, b, c, d] \mapsto (T_l , \Lambda_l) = (\omega^{\frac{1}{2}}\frac{bc}{ad} , \frac{d^N}{a^N}) \ ,
\elea(Whye)
((4), (9), (11) in \cite{R92})\footnote{The variables $(T_r , \Lambda_r), (T_l, \Lambda_l)$ here and $(t_r, \lambda), (t_l, \lambda)$ in the equations (9) (11) of \cite{R92} are related by $(T_r, \Lambda_r) = (t_r^{-1},   \frac{{\rm i} k \lambda}{1 - k^\prime \lambda})$, $(T_l, \Lambda_l) = (\omega^{\frac{1}{2}} t_l ,   \frac{-k \lambda}{k^\prime -  \lambda})$.}.
The hyperelliptic curve  $W_\kappa$ can also be represented in the following different form \cite{R92, R98}. By quotients of symmetries of $W_\kappa$, 
one has the following commutative diagram of Riemann surfaces:
$$
\begin{array}{lllll}
& W_\kappa & \stackrel{\Psi}{\longrightarrow} & \PZ^1 & = W_\kappa/ \langle \theta \rangle \\ [1mm]
& \downarrow \Pi & & \downarrow \pi & \\
W_\kappa /\langle \sigma \rangle=& \PZ^1 &\stackrel{\psi}{\longrightarrow} & 
\PZ^1 & = W_\kappa/ \langle \theta, \sigma \rangle 
\end{array}
$$
where $\Psi , \psi, \Pi , \pi$ are the natural projections with the coordinate expressions: 
$$
\Psi ( T , \Lambda ) = \lambda \ , \ \ \Pi ( T , \Lambda ) = T \ , \ \ 
\psi ( T ) = T^N \ , \ \ \ \ \pi ( \Lambda ) = 
\frac{(1-\kappa \Lambda ) ( 1 - \kappa \Lambda^{-1} )
}{1-\kappa^2} \ \  . 
$$
The $(T, \Lambda)$-coordinates in $W_\kappa$ of branch points for the projections $\Psi$ and $\Pi$ are given by
$$
\begin{array}{ll}
\mbox{Branch \ points \ of \ } \Psi : &
 (\infty ,  0 ), \  ( \infty , \infty ) , \  ( 0 , \kappa ) , \ (0 , \kappa^{-1} ) \ , 
\\ [2mm]
\mbox{Branch \ points \ of \ } \Pi :& ( \omega^{-j} \sqrt[N]{\frac{1+\kappa}{1-\kappa}} , -1 ) , \ \ (\omega^{-j}  \sqrt[N]{\frac{1-\kappa}{1+\kappa}} , 1) \ \ 
\ , \ 1 \leq j \leq N \ ,
\end{array}
$$
where $ \sqrt[N]{\frac{1-\kappa}{1+\kappa}} := \sqrt[N]{|\frac{1-\kappa}{1+\kappa}|}
e^{\frac{\rm i}{N}{\rm arg} (\frac{1-\kappa}{1+\kappa})}$.  Using the birational transformations, 
$$
w = \frac{\kappa}{1- \kappa^2} ( \Lambda - \frac{1}{\Lambda} ) \ , \ \, \ \, \
\Lambda = \frac{1}{2\kappa} \{ (1-\kappa^2)( w - T^N) + \kappa^2 + 1 \} \ ,
$$
one obtains the equivalent form of the curve $W_\kappa$ in terms of $(w, T)$-variables,
$$
W_\kappa \ :  \  w^2  =  ( T^N - \frac{1- \kappa}{1+\kappa} )( T^N - \frac{1+ \kappa}{1-\kappa} ) \ ,  \ \ ( T , w ) \in \CZ^2  \  .
$$
Now we are able to determine all the symmetries of the rapidity curve ${\goth W}$ and its group structure through the hyperelliptic curve $W_{{\rm i}k^\prime/k}$.
By (\req(CPHE)) and (\req(Whye)), ${\goth W}$ is an unramified cover over $W_{{\rm i}k^\prime/k}$ with the $\ZZ_N^2$-covering group $\langle M^{(1)}, M^{(2)} \rangle$ via the map,
$$
\xi : \ {\goth W} \longrightarrow W_{{\rm i}k^\prime/k} \ , \, \ \, \ [a, b, c, d] \mapsto (T , \Lambda ) = (\frac{ac}{bd}, \frac{{\rm i} d^N}{ b^N} )  \ ,
$$
by which the symmetries of $W_{{\rm i}k^\prime/k}$ can be lifted to automorphisms of ${\goth W}$ in the following manner:
\be
M^{(3)} \ \,  \rightarrow \ \,  \sigma \ ; \ \, 
M^{(4)} \ \, \rightarrow \ \, \theta^{-1} \ ; \ \,
R \ \, \rightarrow \ \, \iota \cdot \theta \ ; \ \, M^{(5)} \ \, \rightarrow \ \, \iota
\cdot \theta \cdot \sigma \ .
\ele(lift) 
\begin{proposition} \label{prop:AutW} 
For $N \geq 3$, the automorphism group ${\rm Aut}({\goth W})$ of ${\goth W}$ is generated by $M^{(j)}$, $1 \leq j \leq 5$, and we have the following exact sequence of groups: 
$$
1 \ \longrightarrow \ \ZZ_N^2 \ \longrightarrow \ {\rm Aut}({\goth W}) \ \longrightarrow \ZZ_2 \times D_N \ \longrightarrow \ 1 \ .
$$
As a consequence, the order of ${\rm Aut}({\goth W})$ is equal to $4N^3$. 
\end{proposition}
{\it Proof}. When $\frac{k^\prime}{k} \neq \pm 1 $, ${\rm Aut}(W_{{\rm i}k^\prime/k})$ is generated by  $\sigma, \theta, \iota$ with its structure isomorphic  to $ \ZZ_2 \times D_N $ \cite{R98}. As $\sigma, \theta, \iota$ can be lifted to those of ${\goth W}$ via (\req(lift)), one has the surjective group homomorphism from ${\rm Aut}({\goth W})$ onto ${\rm Aut}(W_{{\rm i}k^\prime/k})$ with the kernel isomorphic to $\ZZ_N^2$. Then follows the result. When $\frac{k^\prime}{k} = \pm 1 $, by replacing $W_{{\rm i}k^\prime/k}, M^{(2)}$ by $W_{k^\prime}, M^{(0)}$ respectively, the same argument again gives the conclusion for ${\rm Aut}({\goth W})$.
$\Box$ \par \vspace{.2in} \noindent

\section{The Descended Forms of $\tau^{(2)}T$ and $T\widehat{T}$ relations on the Hyperelliptic Curve} 
In the study of chiral Potts models, for the $T_p(q)$-eigenvalue problem  one reduces the operators on ${\goth W}$ to those over $W_{k^\prime}$ \cite{B90, MR}; while discussions of the order parameter problem of chiral Potts models were conducted by using the curve $W_{{\rm i}k^\prime/k}$ \cite{B1, B2, B3}. In this section, we consider only the formal case, and derive the functional equations on $W_{k^\prime}$ corresponding  to the $\tau^{(2)}T$ and $T\widehat{T}$ relations on ${\goth W}$. By $T_p(q)= T_p(x_q, y_q)$ and $W_{k'} = {\goth W}/\langle M^{(0)}, M^{(1)} \rangle$ in (\req(CPHE)), for the reduction of $T_p(q)$ to an operator on $W_{k'}$, one needs only to examine the effect of $T_p(q)$ when replacing $q$ by $M^{(0)}(q)$. The relation is given by formula (2.40) in \cite{BBP}:
\be
T_p ( \omega x_q , \omega^{-1} y_q ) = \bigg( \frac{(y_p - \omega x_q)(y_p - \omega^{-1} y_q)}{\mu_p^2 (\omega x_p - y_q) (x_p - x_q )} \bigg)^L X^{-1} T_p (x_q, y_q) \ .
\ele(TM0)
In order to eliminate the scalar-factor in the above right hand side, a procedure of normalizing $T_p (x_q, y_q)$ was given in \cite{B90} via the function $g_p(q)\overline{g}_p(q)$, where $g_p, \overline{g}_p$ are functions on ${\goth W}$ defined by 
$$
\begin{array}{lll}
g_p(q) := & \prod_{n=0}^{N-1} W_{pq}(n) &( = (\frac{\mu_p}{\mu_q})^{\frac{(N-1)N}{2}} \prod_{j=1}^{N-1} (\frac{y_q - \omega^j x_p}{y_p - \omega^j x_q})^{N-j} ) \ , \\
\overline{g}_p(q) := & {\rm det}_N(\overline{W}_{p q}(i-j)) & ( = N^{\frac{N}{2}} e^{\frac{\pi {\rm i}(N-1)(N-2)}{12}} \prod_{j=1}^{N-1} \frac{(t_p - \omega^j t_q)^j}{(x_p - \omega^j x_q)^j (y_p - \omega^j y_q)^j } \ , \ \ {\rm by \ (2.44) \ in \ \cite{BBP}}) \ .
\end{array}
$$
One has
\be
g_p(q)\overline{g}_p(q) = N^{\frac{N}{2}} e^{\frac{\pi {\rm i}(N-1)(N-2)}{12}} (\frac{\mu_p}{\mu_q})^{\frac{(N-1)N}{2}}    \prod_{k=1}^{N-1} \frac{(x_p - \omega^{k} y_q)^k(t_p - \omega^k t_q)^k}{(x_q - \omega^k y_p)^k (x_p - \omega^k x_q)^k (y_p - \omega^k y_q)^k } \ .
\ele(gg-)
By $\mu_p^N(x_p^N-x_q^N)(x_p^N-y_q^N) = \mu_p^{-N}(y_p^N-x_q^N)(y_p^N-y_q^N)$, one obtains the following relation for the function $g\overline{g}$ when changing the variable $q$ to $M^{(0)}(q)$:
\be
g_p(M^{(0)}q)\overline{g}_p(M^{(0)}q) = (-1)^{N-1}  \bigg(\frac{(y_p - \omega x_q)(y_p - \omega^{-1} y_q)}{\mu_p^2 (\omega x_p- y_q)(x_p - x_q)  }\bigg)^N g_p(q) \overline{g}_p(q) \ .
\ele(gg)
By comparing the factors in (\req(TM0)) and (\req(gg)), one leads to the operator
\be
V_p(q) = S_R^{\frac{-1}{2}}T_p(q)/ (g_p(q) \overline{g}_p(q))^{\frac{L}{N}} ,
\ele(Vdef)
with  the size $L$ being only {\it even for N even}\footnote{This requirement was not put in \cite{B90}, but we add it here out of consideration of the factor $(-1)^{N-1}$ in the relation (\req(gg)).}. 
By $X^N=1$ and the relation between ${\goth W}$ and $W_{k'}$, the operator $V_p(q)^N$ depends only on the values of $t_q$ and $ \lambda_q$, hence is defined on the curve $W_{k'}$. Up to $N$th roots of unity, we may write $V_p(q)$ simply by $V_p(t_q, \lambda_q)$. With the same argument, one can see that $\prod_{j=0}^{N-1} V_p(\omega^j t_q, \lambda_q)$ depends on the values of $t_q^N, \lambda_q$, hence becomes a function of the variable $\lambda_q$ only. Indeed, by examining poles of the function, Baxter obtained its precise form ((4) of \cite{B90}):
\be
\prod_{j=0}^{N-1} V_p(\omega^j t_q, \lambda_q) = \eta^L \lambda_q^{\frac{-(N-1)L}{2}} \alpha_p(\lambda_q)^{\frac{-(N-1)}{2}} S(\lambda_q) \ , \ \ \ \ \eta:= e^{\frac{\pi {\rm i}(N-1)(N+4)}{12}} \ ,
\ele(VS)
where $\alpha_p(\lambda)$ is given by (\req(alpha)), and $S(\lambda)$ is a polynomial of $\lambda$ of degree $(N-1)L$. We shall use the operator $V_p( t_q, \lambda_q)$ on $W_{k'}$ to describe the $\tau^{(2)}T$ and $T \widehat{T}$ relations. 

The $T\widehat{T}$ relations on $W_{k'}$ were already given in \cite{B90} as equation (8) there. For the self-contained nature of this article, we represent here a little more detailed derivation on the formula by using (\req(TThat)).
By (\req(TM0)), one has 
$$
\widehat{T}_p ( \omega^j y_q , x_q ) = \bigg(\frac{1}{\mu_p^{2j}} \prod_{k=1}^j  \frac{(y_p - \omega^k y_q)(y_p - \omega^{j-k} x_q)}{(\omega x_p - \omega^{j+1-k} x_q)(x_p - \omega^{k-1} y_q)}\bigg)^L X^{-j} \widehat{T}_p(y_q , \omega^j x_q ) \ ,
$$
by which (\req(TThat)) can be converted into the following form:
$$
\begin{array}{ll}
T_p(x_q, y_q) \widehat{T}_p ( \omega^j y_q , x_q ) =&  r_{p, q} h_{j ; p, q} \bigg( \prod_{k=1}^j  \frac{ (y_p - \omega^k y_q)(y_p - \omega^{j-k} x_q)}{\mu_p^2(\omega x_p - \omega^{j+1-k} x_q)(x_p - \omega^{k-1} y_q)}\bigg)^L  \times \\
& \bigg(X^{-j}  \tau^{(j)}_p(t_q) + \frac{\prod_{k=1}^{j}z(\omega^{k-1} t_q)}{\alpha_p (\lambda_q)} \tau_p^{(N-j)} (\omega^j t_q) \bigg) \ .
\end{array}
$$
By (\req(gg-)), one can derive the identity:
\begin{eqnarray}
&g_p(q)\overline{g}_p(q)g_p(U^{j+1}R^{-1} q)\overline{g}_p(U^{j+1}R^{-1} q)= \nonumber \\  &\frac{N^N \lambda_p^{N-1}(x_p^N-y_q^N)^j(y_p-y_q)^N(y_p- x_q)^N}{e^{\frac{\pi {\rm i}(N-1)(N-2)}{6}} (y_p^N-y_q^N)^{N+j}(y_p^N-x_q^N)^N} ( \prod_{k=1}^j \frac{(y_p - \omega^k y_q)^N}{(x_p-\omega^{k-1}y_q)^N(t_p - \omega^{k-1} t_q)^N} )
(t_p^N-t_q^N)^j \prod_{k=1}^{N-1} (t_p - \omega^{k+j} t_q)^{2k} \ . \nonumber 
\end{eqnarray}
Then by the relation
\begin{eqnarray}
&(r_{p, q} h_{j ; p, q})^{\frac{1}{L}}\prod_{k=1}^j  \frac{ (y_p - \omega^k y_q)(y_p - \omega^{j-k} x_q)}{\mu_p^2(\omega x_p - \omega^{j+1-k} x_q)(x_p - \omega^{k-1} y_q)}  = \frac{N y_p^{2j-2}(y_p -  x_q)(y_p - y_q) (t_p^N-t_q^N)}{\omega^j \mu_p^{2j}(x_p^N - x_q^N) (y_p^N - y_q^N)}  \prod_{k=1}^j  \frac{(y_p - \omega^k y_q)}{(x_p - \omega^{k-1} y_q)(t_p - \omega^{k-1} t_q)} \ , \nonumber
\end{eqnarray}
one has
\begin{eqnarray}
&\frac{(r_{p, q} h_{j ; p, q})^{\frac{N}{L}}(\prod_{k=1}^j  \frac{ (y_p - \omega^k y_q)(y_p - \omega^{j-k} x_q)}{\mu_p^2(\omega x_p - \omega^{j+1-k} x_q)(x_p - \omega^{k-1} y_q)})^N}{g_p(q)\overline{g}_p(q)g_p(U^{j+1}R^{-1} q)\overline{g}_p(U^{j+1}R^{-1} q)} = 
 \frac{\eta^2}{ \prod_{k=1}^{N-1} [ \omega \mu_p^2 y_p^{-4} (t_p - \omega^{j+k} t_q)^2 ]^k} 
(\frac{(y_p^N-x_q^N)(t_p^N-t_q^N)}{y_p^{2N}(x_p^N -x_q^N)})^{N-j} \ . \nonumber
\end{eqnarray}
By which, the $T\widehat{T}$ relations (\req(TThat)) becomes equation (8) of \cite{B90} in  variables $(t, \lambda)$, 
$$
\begin{array}{l}
\alpha_p (\lambda)^{\frac{j}{N}} \zeta(\omega^j t)  V_p(t, \lambda) V_p ( \omega^j t , 
\lambda^{-1} ) = \alpha_p (\lambda) X^{-j}  \tau^{(j)}_p(t) + (\prod_{k=1}^j z(\omega^{k-1} t)) 
\tau_p^{(N-j)} (\omega^j t)  \ , 
\end{array}
$$
where $ \zeta (t):=  \eta^{\frac{-2L}{N}} \prod_{k=1}^{N-1} z (\omega^k t)^{\frac{k}{N}} $. In particular for $j=0$, we have $ \tau_p^{(N)} (t)= \zeta(t)  V_p(t, \lambda) V_p (  t , \lambda^{-1} ) $. Then by (\req(VS)), one arrives formulae (11), (12) in \cite{B90},
\bea(l)
S(\lambda) S(\lambda^{-1}) = \tau_p^{(N)} (t) \tau_p^{(N)} (\omega t) \cdots \tau_p^{(N)} (\omega^{N-1}t) \ , \\
V_p (t, \lambda)^N = \frac{\eta^L}{ \lambda^{\frac{(N-1)L}{2}} \alpha_p(\lambda )^{N} S(\lambda^{-1})}\prod_{j=1}^N \bigg( \alpha_p (\lambda) X^{-j}  \tau^{(j)}_p(t) + (\prod_{k=1}^{j}z(\omega^{k-1} t)) \tau_p^{(N-j)} (\omega^j t) \bigg) \ .
\elea(StauVN)
By the commutativity of operators $X, \tau_p^{(j)} (t), V_p(t, \lambda )$ and $ S(\lambda)$, their eigenvalues again satisfy the relations (\ref{Fus1}), (\ref{Fus2}), (\req(StauVN)), regarded as scalar functions on $W_{k'}$. By which, one can solve first $\tau_p^{(j)}(t)$ by (\ref{Fus1}), (\ref{Fus2}), then by (\req(StauVN)) obtain $S(\lambda)$, hence eigenvalues of $V_p(t, \lambda )$, (equivalently, those of $T_p(q)$). All the above relations should in principle place one well on road to solving the eigenvalue problem of chiral Potts model; 
however as $(t, \lambda)$ is the "coordinates" of a higher genus curve $W_{k'}$, it is still a difficult problem to extract explicit solutions for a finite site $L$. Nevertheless, one can use these equations to obtain the maximum eigenvalues \cite{B90} in the thermodynamic limit as $L$ tends to $\infty$, as well as in the discussion of excitation spectrum in \cite{MR}.

We now identify the $(t, \lambda)$-form of $\tau^{(2)}T$ relation. By (\req(gg-)), the relation (\req(tauT)) becomes  
$$
\begin{array}{ll}
&y_p^{2L}( \prod_{k=1}^{N-1} \frac{(t_p - \omega^{k+1} t_q)^k}{(\omega x_q - \omega^k y_p)^k (x_p - \omega^{k+1} x_q)^k  }
)^{\frac{L}{N}}\tau^{(2)}_p(t_q) V_p(\omega t_q, \lambda_q)  \\
= &(-\omega )^L \bigg(\frac{( x_q -\omega^{-1} y_p )^N (t_p -t_q)^N }{(x_p - x_q)^N } \prod_{k=1}^{N-1} \frac{(t_p - \omega^k t_q)^k}{(x_q - \omega^k y_p)^k (x_p - \omega^k x_q)^k  }\bigg)^{\frac{L}{N}} V_p(t_q, \lambda_q) + \\ 
&(-\omega )^{-L} \bigg(\frac{ \lambda_p^2(x_p -  \omega x_q)^N (t_p - \omega t_q)^N}{ ( x_q - \omega^{-2} y_p )^N } \prod_{k=1}^{N-1} \frac{(t_p - \omega^{k+2} t_q)^k}{(\omega^2 x_q - \omega^k y_p)^k (x_p - \omega^{k+2} x_q)^k  }
\bigg)^{\frac{L}{N}} X V_p(\omega^2 t_q , \lambda_q ) \ , 
\end{array}
$$
hence one has
$$
\begin{array}{ll}
y_p^{2L}\tau^{(2)}_p(t_q) V_p(\omega t_q, \lambda_q) =  & 
\bigg(\frac{(-1)^N\omega^{N(N+1)/2} (x_q^N -  y_p^N )(t_p^N - t_q^N )   }{(x_p^N - x_q^N) } \bigg)^{\frac{L}{N}} V_p(t_q , \lambda_q ) + \\ &  
\lambda_p^{\frac{2L}{N}}(t_p - \omega  t_q)^{2L} \bigg(\frac{  (x_p^N-x_q^N)
}{ (-1)^N \omega^{N(N+1)/2}( x_q^N -  y_p^N ) (t_p^N - t_q^N ) } 
\bigg)^{\frac{L}{N}}  X V_p(\omega^2 t_q , \lambda_q ) \ . 
\end{array}
$$
By $(-\omega )^N \omega^{N(N-1)/2} = -1$ and $\frac{ (y_p^N  - x_q^N ) }{(x_p^N - x_q^N) } = \frac{  \lambda_p \lambda_q - 1 }{ \lambda_p^{-1} \lambda_q - 1 }$, we obtain the $\tau^{(2)}V$ relation on the variable $(t_q, \lambda_q) \in W_{k'}$ for a fixed $(t_p, \lambda_p) \in W_{k'}$:
\bea(c)
(\frac{1-k'\lambda_p}{k})^{\frac{2L}{N}}\tau^{(2)}_p(t_q) V_p(\omega t_q , \lambda_q) = \\
\bigg(\frac{  (\lambda_p^2 \lambda_q - \lambda_p )(t_p^N -  t_q^N ) }{
  \lambda_q - \lambda_p }  \bigg)^{\frac{L}{N}} V_p( t_q , \lambda_q ) +  
(t_p - \omega t_q)^{2L} \bigg(\frac{ \lambda_p \lambda_q - \lambda_p^2 } 
{  (\lambda_p \lambda_q - 1 )(t_p^N -  t_q^N ) }
\bigg)^{\frac{L}{N}}  X V_p(\omega^2 t_q , \lambda ) \ . 
\elea(tauV)
In particular, in the superintegrable case where $\lambda_p=1, t_p=(\frac{1-k'}{1+k'})^\frac{1}{N}$, the relation (\req(tauV)) becomes
$$
\begin{array}{c}
(\frac{1-k'}{1+k'})^{\frac{L}{N}}\tau^{(2)}_p ( t_q) V_p (\omega t_q , \lambda_q) = \\ 
\bigg( \frac{1-k'}{1+k'} - t_q^N  \bigg)^{\frac{L}{N}} V_p( t_q , \lambda_q ) +  
\bigg((\frac{1-k'}{1+k'})^{\frac{1}{N}} - \omega t_q \bigg)^{2L} \bigg(   \frac{1-k'}{1+k'} - t_q^N  
\bigg)^{\frac{-L}{N}}  X V_p(\omega^2 t_q , \lambda_q ) \ . 
\end{array}
$$
Therefore, if $\tau^{(2)}_p(t_q)$ coupling with a function $V_p(t_q , \lambda_q)$ of $W_{k'}$ forms a solution of the above relation, $V_p(t_q , \lambda_q^{-1})$ is also a solution with the same $\tau^{(2)}_p( t_q)$. By the procedure of solving the eigenvalue $V_p(t_q , \lambda_q)$ in chiral Potts model, $V_p(t_q , \lambda_q)$ is not invariant under the change of $\lambda_q$ to $\lambda_q^{-1}$. Hence the correspondence from the $V_p$-eigenvalues to $\tau^{(2)}_p$-eigenvalues is at least a 2-1 map. Therefore the degeneracy of  $\tau^{(2)}_p$-eigenvalues occurs when $p$ is the superintegrable element, an analogy to the discussion of $T$-$Q_{72}$ relation for the eight-vertex model in \cite{FM}.

\section{Concluding Remarks}
In this paper, we  made a clear mathematical derivation on the descending process from the six-vertex model to the chiral Potts $N$-state model following the works \cite{BBP, BazS}.
We start with the Yang-Baxter solution (\req(G)) of six-vertex model carrying an arbitrary four-vector ratios $p$, then reinterpret the descendant relation of six-vertex model and chiral Potts model through the fusion relations (\ref{Fus1}) (\ref{Fus2}) of $\tau^{(j)}_p$, finally reach the chiral Potts constraint (\ref{rapidC}) for the rapidity $p$. 
The finding does suggest that studies of all $\tau^{(j)}$-families should be important for understanding the mathematics in the chiral Potts transfer matrices. Although the operators $\tau^{(j)}$ in statistical mechanics are in many respects well understood physically, the mathematical investigation on the fusion relations about these operators still lags behind. Certain interesting topics are expected to arise by exploring deeper into their mathematical structures. From the relations between the rapidity curve ${\goth W}$ and  three genus-$(N-1)$ hyperelliptic curves with $D_N$-symmetry, we determined the structure of ${\rm Aut}({\goth W})$, hence all the symmetries of rapidities. Through one of these hyperelliptic curves, $W_{k'}$ in (\req(Wk')), we obtain the reduced form (\req(tauV)) of $\tau^{(2)}T$ relation on $W_{k'}$, which is descended from ${\goth W}$. By which, we are able to indicate the degeneracy of $\tau^{(2)}_p$-eigenvalues when $p$ is a superintegrable point, a similar phenomenon for $T$-$Q_{72}$ relation  discussion of eight-vertex model in \cite{FM}. The comparison is also one of the motivations for our investigation in this article. Further developments along this line are now under consideration, and progress is expected.

\section*{Acknowledgements}
I wish to take this opportunity to thank R. Baxter for his stimulating lectures on topics related to chiral Potts models during his visit of Taiwan in November 2001, which contributed a great deal for the author to understand the subjects in this article.
This work has been supported by NSC 92-2115-M-001-023.

\end{document}